\documentclass[12pt,letterpaper]{article}    
\usepackage{osajnl2} 
\usepackage{amsfonts}      
\usepackage{amsmath}        
\usepackage{amssymb}        
\usepackage{amsbsy}         
\usepackage[mathscr]{eucal} 
\usepackage{graphicx}       

\begin{document}
\title{Normal mode coupling and Bloch states in elliptically birefringent 1D photonic crystals}
\author{Amir A Jalali and Miguel Levy}
\address{Michigan Technological University, Department of Physics, 1400 Townsend Dr, Houghton, MI 49931, USA}
\begin{abstract}
An analysis is presented of frequency versus wave-vector dispersion in elliptically birefringent one-dimensional layered periodic structures.  The presence of local normal mode polarization state variations from one layer to the next is found to lead to mode coupling and to affect the wave-vector dispersion and the character of the Bloch states of the system.
\end{abstract}
\section{Introduction}
A number of publications have discussed the propagation of electromagnetic waves in one-dimensional birefringent stratified media~\cite{Yeh1979,Mandatori2003}. Of particular interest is the formulation pioneered by P. Yeh that employs a translation matrix approach for periodic structures.  The propagation of light across a single period is analyzed in conjunction with Floquet's theorem to determine the dispersion relation and Bloch waves for the system.  This approach has been used to study the properties of layered media where the optic axes are rotated from one layer to the next~\cite{Mandatori2003}. Numerical solutions have been found for normal incidence of the light for identical birefringent plates with alternating azimuth angles and layer thicknesses~\cite{Yeh1979}.

More recently the present authors have used this technique to study elliptically birefringent media~\cite{Levy2007}, where elliptically polarized normal modes characterize the system locally.  Analytic solutions were found for light normally incident into stratified media consisting of alternating magneto-optic layers with different gyration vectors~\cite{Landau1984} and birefringence levels~\cite{Levy2007}. Adjacent layers were assumed to have their anisotropy axes aligned to each other.  This model captures some important features of one-dimensional magnetophotonic crystal waveguides, particularly the presence of magneto-optical gyrotropy and alternating birefringence levels.  Such systems are presently being studied for use in integrated fast optical switches and ultra-small optical isolators~\cite{Levy2006a,Li2005,Li2005a}.  Work on these systems extends prior theoretical and experimental efforts on magnetophotonic crystals in order to encompass distinctive properties of planar structures~\cite{Khanikaev2005,Figotin2001,Inoue1999}.

A particularly interesting feature of periodic elliptically birefringent gyrotropic media concerns the character of Bloch mode and dispersion branch solutions to the Floquet theorem.  The purpose of the present article is to discuss this facet of the problem because it contains  interesting implications.

The work presented here traces the origin of normal mode coupling to the simultaneous presence of gyrotropy and linear birefringence in the periodic system, as normal modes change from one layer to the next. As a result, a new kind of parameter appears that couples layer-dependent normal modes to produce new Bloch states and propagation constants. This coupling depends on the relative parameters of adjacent layers but not on the precise positioning of the origin.  We present an approximate analytical expression for the dispersion relation.

 The present paper deals with elliptically birefringent media having aligned anisotropy axes. It contains three sections after this introduction. Section 2 presents the formalism we will use to discuss the problem, and section 3 discusses the main results that follow from this formalism. The last section presents a series of concluding remarks.

\section{Waves in a birefringent magnetophotonic medium}
\begin{figure}[t]
\includegraphics{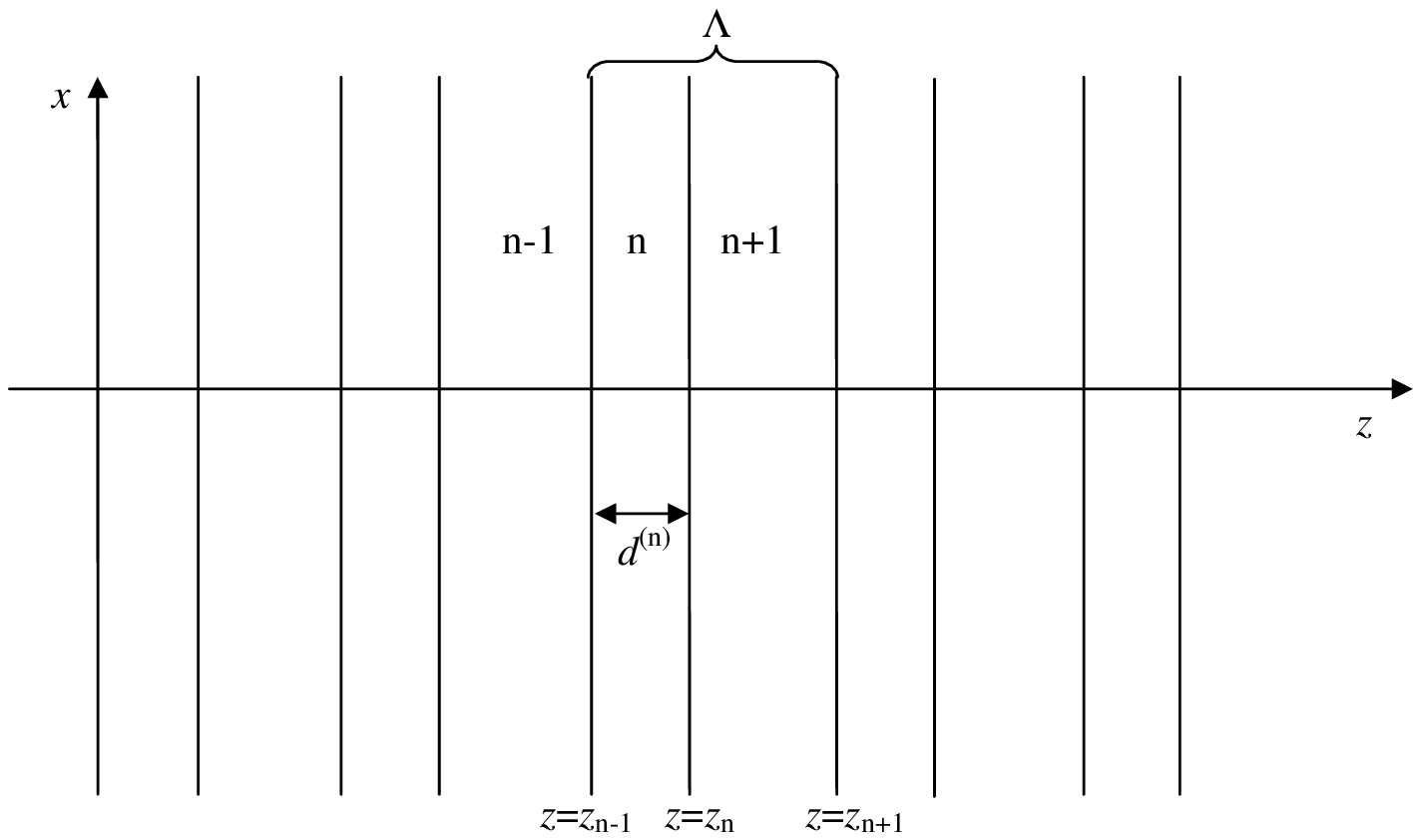}
\caption{Schematic diagram of a one-dimensional birefringent magnetophotonic crystal with period of $\Lambda$. The magnetophotonic crystal extend indefinitely in the $x$ and $y$ directions. A plane wave is incident normally to the layered structure. }
\label{fig1}
\end{figure}

In the optical wavelength regime, the permeability of a birefringent uniaxial magnetooptic medium is very close to the permeability of vacuum $\mu_0$, its relative permeability close to unity. The relative permittivity tensor $\tilde{\epsilon}$ of the medium for magnetization along the $z-$axis, has the form~\cite{Wolfe1988}
\begin{equation}
\label{1}
    \tilde{\epsilon}=\left(
    \begin{array}{c c c}
    \epsilon_{xx} & i\epsilon_{xy} & 0\\
    -i\epsilon_{xy} & \epsilon_{yy} & 0\\
    0 & 0 & \epsilon_{zz}\\
    \end{array}
    \right),
\end{equation}
where we assume no absorption of the light in the medium. This implies that all components of the relative permittivity ($\epsilon_{i,j}, i,j=x,y,z$) are real and it is not assumed that $\epsilon_{xx}=\epsilon_{yy}$. By solving the wave equation upon normal incidence of a monochromatic plane wave propagating parallel to the $z$ axis (Fig. \ref{fig1}) on a birefringent magnetooptic medium, one obtains eigenmodes
\begin{eqnarray}
 \label{3}
    \mathbf{\hat{e}}_{\pm}=\frac{1}{\sqrt{2}}
    \left(
                \begin{array}{c}
                \cos\alpha \pm \sin \alpha \\
                \pm i\cos \alpha - i \sin \alpha\\
                0\\
                \end{array}
    \right),
\end{eqnarray}
with refractive indices $n_{\pm}$, $n_{\pm}^2 = \bar{\epsilon} \pm  \sqrt{\Delta^2+\epsilon_{xy}^2}$.
Here $\bar{\epsilon}=(\epsilon_{yy} + \epsilon_{xx})/2$, $\Delta~=~(\epsilon_{yy}-~\epsilon_{xx})/2$. The parameter $\alpha$ referred hereafter as \emph{elliptical birefringence parameter}, is given by $\tan (2\alpha)=\Delta/\epsilon_{xy}$

\section{One-dimensional birefringent magnetophotonic crystals}
Consider a plane wave normally incident on a periodic stack structure of alternating elliptically  birefringent magnetophotonic layers (Fig.~\ref{fig1}), where the elliptical birefringence parameters of adjacent layers may differ. The Bloch states for this system satisfy the Floquet-Bloch theorem through the following eigenvalue equation.
\begin{eqnarray}
\label{13}
  \mathbf{T}^{(n-1,n+1)}\mathbf{E} &=& \lambda \mathbf{E},
\end{eqnarray}
where the transfer matrix $\mathbf{T}^{(n-1,n+1)}$  relates the four eigenmode amplitudes $\mathbf{E}$ in the first layer of a unit cell to the amplitudes in the first layer of the adjacent unit cell and $\lambda$ is a scalar quantity \cite{comment}. Upon formulation of the problem in terms of local normal modes $\mathbf{\hat{e}}^{(n)}_{\pm}$ and $\mathbf{\hat{e}}^{(n+1)}_{\pm}$ the above eigenvalue problem can alternatively be expressed as illustrated in~\cite{Levy2007},
 \begin{eqnarray}
 \label{eig}
    \mathbf{S}^{(n)}_{c} \mathcal{E}&=& \lambda \mathbf{B}^{(n,n+1)} \mathcal{E},
 \end{eqnarray}
 with $\mathbf{S}^{(n)}_{c}$, $\mathbf{B}^{(n,n+1)}$, and $\mathcal{E}$  given by,~\cite{Levy2007}
 \begin{eqnarray}
S^{(n)}_{c}=
\begin{pmatrix}
  \cos \beta^{(n)}_+ & \frac{i}{n^{(n)}_+}\sin \beta^{(n)}_+ & 0 & 0 \\
  i n^{(n)}_+\sin \beta^{(n)}_+  & \cos \beta^{(n)}_+ & 0 & 0 \\
  0 & 0 & \cos \beta^{(n)}_- & \frac{i}{n^{(n)}_{-}}\sin \beta^{(n)}_{-} \\
  0 & 0 & i n^{(n)}_{-}\sin \beta^{(n)}_{-} & \cos \beta^{(n)}_{-} \\
\end{pmatrix},
\end{eqnarray}
\begin{eqnarray}
B_{1,1}&=& B_{2,2}=\cos^{2}\chi^{(n,n+1)} \cos{\beta^{(n+1)}_{+}} +  \sin^{2}\chi^{(n,n+1)} \cos{\beta^{(n+1)}_{-}},\\
B_{1,2}&=&-i \frac{\sin^{2}\chi^{(n,n+1)}  \sin{\beta^{(n+1)}_{-}}}{n^{(n+1)}_{-}}-i \frac{\cos^{2}\chi^{(n,n+1)} \sin{{\beta^{(n+1)}_{+}}}}{n_{+}^{(n+1)}},\\
B_{1,3}&=&B_{3,1}=B_{2,4}=B_{4,2}= \frac{1}{2}\sin2\chi^{(n,n+1)}\left(\cos{{\beta^{(n+1)}_{-}}}-\cos{{\beta^{(n+1)}_{+}}}\right), \\
B_{1,4}&=&B_{3,2}=\frac{i \sin2\chi^{(n,n+1)}}{2 \;n^{(n+1)}_{-} n^{(n+1)}_{+}} \left(n^{(n+1)}_{-} \sin \beta^{(n+1)}_{+} - n^{(n+1)}_{+} \sin \beta^{(n+1)}_{-} \right),\\
B_{2,1}&=& -i n^{(n+1)}_{-} \sin^{2}\chi^{(n,n+1)} \sin{\beta^{(n+1)}_{-}}-i n_{+}^{(n+1)}\cos^{2}\chi^{(n,n+1)} \sin{{\beta^{(n+1)}_{+}}}, \\
B_{2,3}&=&B_{4,1}= i\; \frac{1}{2}\sin2\chi^{(n,n+1)}\left(n^{(n+1)}_{+} \sin \beta^{(n+1)}_{+} -  n^{(n+1)}_{-} \sin \beta^{(n+1)}_{-}\right),\\
B_{3,3}&=&B_{4,4}= \cos^{2}\chi^{(n,n+1)}  \cos{\beta^{(n+1)}_{-}} + \sin^{2}\chi^{(n,n+1)}  \cos{\beta^{(n+1)}_{+}},\\
B_{3,4} &=& -i \frac{\cos^{2}\chi^{(n,n+1)}  \sin{\beta^{(n+1)}_{-}}}{n^{(n+1)}_{-}}-i \frac{\sin^{2}\chi^{(n,n+1)} \sin{{\beta^{(n+1)}_{+}}}}{n_{+}^{(n+1)}},\\
B_{4,3}&=& -i n^{(n+1)}_{-} \cos^{2}\chi^{(n,n+1)}  \sin{\beta^{(n+1)}_{-}}-i n_{+}^{(n+1)} \sin^{2}\chi^{(n,n+1)} \sin{{\beta^{(n+1)}_{+}}},
\end{eqnarray}
 and
 \begin{eqnarray}
 \mathcal{E}=\mathbf{\Phi}^{(n,n+1)}\mathbf{P}^{(n+1)} \mathbf{E}
 \end{eqnarray}
where $\beta^{(n)}_{\pm}=(\omega/c) n^{(n)}_{\pm} d^{(n)}$, $\chi^{(n,n+1)}=\alpha^{(n)}-\alpha^{(n+1)}$, $\mathbf{P}^{(n+1)}$ is a propagation matrix through layer $n+1$,
\begin{eqnarray}
\nonumber \mathbf{\Phi}^{(n,n+1)}&=&\\
  && \nonumber \begin{pmatrix}
    \cos \chi^{(n,n+1)} & \cos \chi^{(n,n+1)} & \sin \chi^{(n,n+1)} & \sin \chi^{(n,n+1)} \\
    n^{(n+1)}_{+} \cos \chi^{(n,n+1)} & -n^{(n+1)}_{+} \cos \chi^{(n,n+1)} & n^{(n+1)}_{-} \sin \chi^{(n,n+1)} & -n^{(n+1)}_{-} \sin \chi^{(n,n+1)} \\
    -\sin \chi^{(n,n+1)} & -\sin \chi^{(n,n+1)} & \cos \chi^{(n,n+1)} & \cos \chi^{(n,n+1)} \\
     -n^{(n+1)}_{+} \sin \chi^{(n,n+1)} & n^{(n+1)}_{+} \sin \chi^{(n,n+1)} & n^{(n+1)}_{-} \cos \chi^{(n,n+1)} & -n^{(n+1)}_{-} \cos \chi^{(n,n+1)} \\
  \end{pmatrix},\\
\end{eqnarray}
and
\begin{eqnarray}
    \mathbf{E}^{(n)}&=&\left(E^{(n)}_{01} e^{-i\frac{\omega}{c}n^{(n)}_{+}(z-z_n)}+ E^{(n)}_{02} e^{i\frac{\omega}{c}n^{(n)}_{+}(z-z_n)}\right)\;\mathbf{\hat{e}_{+}}^{(n)}\\
              \nonumber &&+\left(E^{(n)}_{03} e^{-i\frac{\omega}{c}n^{(n)}_{-}(z-z_n)}+ E^{(n)}_{04} e^{i\frac{\omega}{c}n^{(n)}_{-}(z-z_n)}\right)\;\mathbf{\hat{e}_{-}}^{(n)}.
\end{eqnarray}
Using the fact that for a wide range of materials the off-diagonal elements in the relative permittivity tensor and the linear birefringence $\Delta$ are much smaller than the diagonal elements, we expand all elements in the $\mathbf{B}^{(n,n+1)}$ matrix up to first order in $\Delta n=(1/2)(n_+ - n_-)$ and solve the resultant eigenvalue equation to obtain a dispersion relation given up to first order in $\Delta n$ by
    \begin{eqnarray}
  \label{647}
    \cos K_{\pm} \Lambda & \approx & \left(\cos{\beta^{(n+1)}_{\pm}} \cos{\beta^{(n)}_{\pm}}-\frac{1}{2} N_{\pm} \sin \beta^{(n+1)}_{\pm}\sin \beta^{(n)}_{\pm}\right) \\
    \nonumber && \pm 2\delta^{(n+1)} \sin 2 \bar \beta_{0} \sin^2 \chi^{(n,n+1)}.
  \end{eqnarray}
Here $\bar \beta_0=(1/2)(\beta^{(n)}_{0}+\beta^{(n+1)}_{0})$, $\delta^{(n)}=(\omega/c)\Delta n^{(n)} d^{(n)}$, and
\begin{eqnarray}
N_{\pm} &=& \frac{n^{(n)}_{\pm}}{n^{(n+1)}_{\pm}} + \frac{n^{(n+1)}_{\pm}}{n^{(n)}_{\pm}},
\end{eqnarray}
with $\beta^{(n)}_{0}=\omega/c (\bar n^{(n)} d^{(n)})$ and $\bar n^{(n)}=(1/2)(n^{(n)}_{+}+n^{(n)}_{-})$.

The Bloch eigenmodes are given by
\begin{eqnarray}
 \label{21}
 \mathbf{E}^{(n)}=\left(\mathbf{\Phi}^{(n,n+1)}\mathbf{P}^{(n+1)}\right)^{-1}
    \begin{pmatrix}
     \frac{\frac{-i}{n^{(n)}_{+}}\sin \beta^{(n)}_{+}+B_{1,2}e^{\pm i K_{\pm} \Lambda} }{\cos \beta^{(n)}_{+}-B_{1,1}e^{\pm i K_{\pm} \Lambda}} \\
     1\\
     \frac{\frac{-i}{n^{(n)}_{-}}\sin \beta^{(n)}_{-}+B_{3,4}e^{\pm i K_{\pm} \Lambda }}{\cos \beta^{(n)}_{-}-B_{3,3}e^{\pm i K_{\pm} \Lambda}} \\
     1\\
     \end{pmatrix}.
 \end{eqnarray}
Note that this expression depends on the relative birefringence parameter $\chi^{(n,n+1)}$ through the components of the matrix $\mathbf{B}^{(n,n+1)}$.

To first order in $\Delta n$ we obtain an analytical solution for the dispersion relation, Eq.~(\ref{647}), since the off-block diagonal terms appear as higher order in $\Delta n$ in the secular determinant.  Deviations from the exact dispersion occur in higher than linear order.  Note that there is now a dependence on the coupling parameter $\chi^{(n,n+1)}$, so that the wave-vector is, in fact, affected by this coupling. The condition for $\chi^{(n,n+1)}$ to differ from zero is $\left(\Delta^{(n)}/\epsilon^{(n)}_{xy}\right) \neq \left(\Delta^{(n+1)}/\epsilon^{(n+1)}_{xy}\right) + m \pi$, where $m$ is an integer.
In purely isotropic and purely circularly birefringent one-dimensional photonic crystals $\chi^{(n,n+1)}=0$. In those cases the coupling between normal modes disappears. This is in agreement with the work presented in the Refs.~\cite{Vytovtov2005,Inoue1999} on periodic stratified isotropic media. Approximate solutions using the formalism presented above are found to be in close numerical agreement with the exact solutions for most currently used materials. The character of the Bloch modes is also affected by the coupling parameter $\chi^{(n,n+1)}$. Whereas Bloch states for non-elliptically birefringent gyrotropic one-dimensional stacks are still circularly polarized, elliptically birefringent stacks can have Bloch modes whose polarization states differ from that of the normal modes in each layer, and that depend on $\chi^{(n,n+1)}$ according to Eq.~(\ref{21}).
\section{Conclusions}
Mode coupling as a result of variations in normal mode polarization states in elliptically birefringent one-dimensional periodic structures is reported. Interlayer normal mode coupling in such media affects the polarization of the Bloch mode states and the wave-vector frequency dependence. This interlayer coupling is absent in isotropic and circularly birefringent media. In all such cases the Bloch mode polarization state coincides with that of the individual constituent layers of a bi-layered unit cell system. Elliptically birefringent magneto-optic media with different elliptically birefringent states in adjacent layers, on the other hand, possess Bloch mode states that depend on the coupling between normal modes in different layers.

\section*{Acknowledgments}
This material is based upon work supported by the National Science Foundation under Grant No. ECCS-0520814.

\newpage

%
%
%
%

\begin{thebibliography}{12}
\bibitem{comment}
We use the Greek letter $\lambda$ to represent the eigenvalue of the Bloch equation. It is not used as a symbol for wavelength.
\end{thebibliography}
\end{document}